\documentclass[twocolumn]{jpsj2}
\usepackage{subfigure}
\newcommand{\bS}{\boldsymbol{S}}
\newcommand{\de}{\partial}

\newcommand{\Neel}{\mathrm{N\acute{e}el}}

\def\runauthor{M. {\sc Nakane}, Y. {\sc Fukumoto} and A. {\sc Oguchi}}

\title{Ground-State Phase Diagram of the ${\bf XXZ}$ Model on a Railroad-Trestle Lattice with Asymmetric Leg Interactions}
\author{Masawo {\sc Nakane}, Yoshiyuki {\sc Fukumoto} and Akihide {\sc Oguchi}}
\inst{Department of Physics, Faculty of Science and Technology, Tokyo University of Science, Noda, Chiba 278-8510  }
\recdate{March 10, 2006}
\abst{Using the bosonization and level spectroscopy methods, we study the ground-state phase diagram of a $XXZ$ antiferromagnet
on a railroad-trestle lattice with asymmetric leg interactions. It is shown that the asymmetry does not change
the dimer/N\'{e}el transition line significantly, which agrees with the expectation based on a naive bosonization procedure,
but it does change the dimer/spin-fluid transition line. To understand this observation, we analyze eigenvectors of the ground state,
dimer excitation, doublet excitation and N\'{e}el excitation, and find that only the doublet excitation is affected by the asymmetric interaction.}
\kword{railroad-trestle lattice, sawtooth lattice, XXZ model, exact singlet dimer state, level spectroscopy method, bosonization}
\begin{document}
\maketitle

%%%%%%%%%%%%%%%%%%%%%%%%%%%%%%%%%%%%%%%%%%%%%%%%%%%%%%%%%%%%%%
\section{Introduction}%%%%%%%%%%%%%%%%%%%%%%%%%%%%%%%%%%%%%%%%%%%%%%%%%%%
%%%%%%%%%%%%%%%%%%%%%%%%%%%%%%%%%%%%%%%%%%%%%%%%%%%%%%%%%%%%%%

Spin-1/2 Heisenberg antiferromagnets on a railroad-trestle lattice have been investigated intensively, 
because they show interesting quantum phase transitions and critical phenomena.\cite{haldane1,fukuyama1,fukuyama2,tonegawa1,okamoto}
Haldane\cite{haldane1} and Nakano and Fukuyama\cite{fukuyama1,fukuyama2} discussed the phase diagram by using the bosonization and renormalization group methods.
Tonegawa and Harada numerically estimated the dimer-N\'{e}el transition point.\cite{tonegawa1}
It is, however, known that a quantitative estimation of the critical point is difficult because of  the nature of the Berezinskii-Kosterlitz-Thouless (BKT) transition.
To overcome this difficulty, Nomura and coworkers developed the level spectroscopy (LS) method,
in which the transition point between the dimer and N\'{e}el (spin-fluid (SF)) phases is estimated by the intersection of energies of the dimer and N\'{e}el (doublet) excitations, 
and they succeeded in obtaining a quantitative result of the phase diagram for the $XXZ$ model on a railroad-trestle lattice.\cite{okamoto,nomura0,nomura1,nomura2,hirata}

The sawtooth-lattice Heisenberg antiferromagnet has been studied by several authors because of the following two reasons:
(i) this system is one-dimensional counter-part of the Kagom\'{e} lattice
and shows a two-stage entropy release as a function of temperature,\cite{kubo,Otsuka} 
and (ii) the sawtooth lattice is realized in $\mathrm{YCuO_{2.5}}$.\cite{sen1}
In this paper, we consider a model that contains both the sawtooth and railroad-trestle models as special cases, and study the ground-state phase diagram in a unified way.

\begin{figure}[h]
\begin{center}
\includegraphics[width=0.9\linewidth]{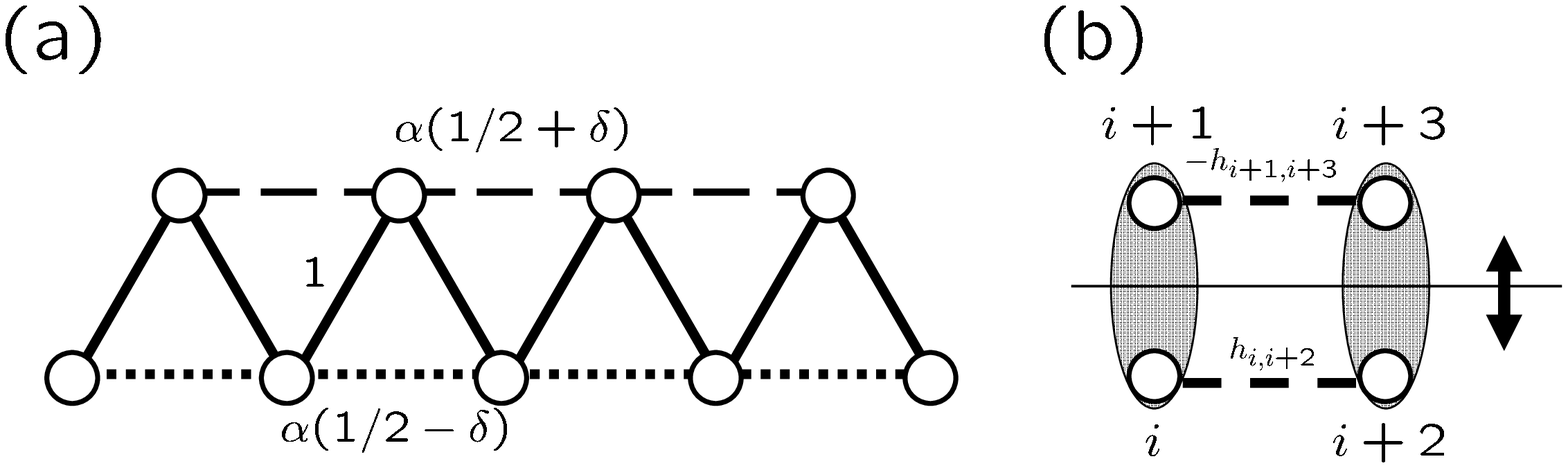}
\end{center}
\caption{Graphical representations of (a) the asymmetric railroad-trestle model and (b) a four spin problem (see text). 
In (b), each dashed line represents $h_{i,i+2}$ or $-h_{i+1,i+3}$, and the thin solid line represents a reflection plane.
The hatched spin pairs are used to define the two-spin basis set.}
\label{fig:model}
\end{figure}

We introduce the model Hamiltonian
\begin{equation}
     H = \sum_{i=1}^L h_{i,i+1} + \alpha\sum_{i=1}^L \left\{ 1/2 + (-1)^i\delta \right\}h_{i,i+2},
\label{eq:1}
\end{equation}
with $h_{i,j} = S^x_iS^x_j + S^y_iS^y_j + \Delta S^z_iS^z_j$. (See Fig.~\ref{fig:model}(a).)
Here, $L$ denotes the total number of spins, and the periodic boundary condition $\bS_{L+1} = \bS_{1}$ is assumed. 
We also confine ourselves to $0 \leq \alpha \leq 1$, $-1 \leq \Delta$ and $0 \le \delta \le 1/2$.
This model is reduced to the railroad-trestle model and sawtooth model, respectively, at $\delta=0$ and $1/2$.\cite{chen1}
The quantitative ground-state phase diagram of the railroad-trestle model with $\Delta \geq 0$ was obtained in ref.~6 for the first time,
and the extension to the $\Delta<0$ region was made in ref.~9, where a direct transition between the dimer state and the ferromagnetic state was found.
Here, we study how the asymmetry parameter $\delta$ affects the phase boundaries.

%----------------------------------------------------
%----------------------------------------------------

This paper is organized as follows: in \S2, we derive some basic properties of the present model. 
We show that the model has an exact dimer singlet ground state at $\alpha=1$, and a naive bosonization procedure predicts that critical lines in the ground-state phase diagram
do not depend on $\delta$.
In \S3, we describe our procedure of numerical calculations based on the LS method.
We present the ground-state phase diagram in the $\alpha-\Delta$ plane in \S4, and discuss the $\delta$-dependence of the transition lines in \S5.
We summarize our results in \S6.

%%%%%%%%%%%%%%%%%%%%%%%%%%%%%%%%%%%%%%%%%%%%%%%%%%%%%%%%%%%%%%
\section{Basic Properties of the Model}%%%%%%%%%%%%%%%%%%%%%%%%%%%%%%%%%%%%%%%%%%
%%%%%%%%%%%%%%%%%%%%%%%%%%%%%%%%%%%%%%%%%%%%%%%%%%%%%%%%%%%%%%

At $\alpha=1$, the system has an exact dimer singlet ground state, independent of $\delta$.
To see this, we first write the singlet dimer state as
\begin{equation}
     [i,j] = \frac{1}{\sqrt{2}}\left\{ \left| \uparrow_i\downarrow_j\right> - \left| \downarrow_i\uparrow_j \right> \right\},
\label{eq:2}
\end{equation}
and define
\begin{subequations}
\begin{eqnarray}
     \Phi_1(L) &&\hspace{-7mm}= [1,2][3,4]\cdots[L-1,L],\label{eq:3a}  \\
     \Phi_2(L) &&\hspace{-7mm}= [2,3][4,5]\cdots[L,1].  \label{eq:3b}
\end{eqnarray}
\end{subequations}
We rewrite eq.~(\ref{eq:1}) with $\alpha=1$ as
\begin{equation}
     H_{\alpha=1} = H_{\alpha=1,\delta=0} + \delta\sum_{i}(-1)^ih_{i,i+2}.
\label{eq:4}
\end{equation}
The first term, $H_{\alpha=1,\delta=0}$, is simply the railroad-trestle model at the Majumdar-Ghosh point, 
and thus, it has doubly degenerate exact dimer singlet ground states, $\Phi_1(L)$ and $\Phi_2(L)$.\cite{majumdar1,majumdar2,Shastry}
In order to show that the dimer singlet state is an eigenstate of eq.~(\ref{eq:4}), it is convenient to use the dimer basis set
and the reflection operation about the thin solid line shown in Fig.~\ref{fig:model}(b).
The two-dimer singlet state $[i,i+1][i+2,i+3]$ has even parity under both the reflection and spin inversion operations.
The operation of $h_{i,i+2} - h_{i+1,i+3}$ to a two-dimer state changes the parity under the reflection operation.
Thus, $(h_{i,i+2} - h_{i+1,i+3})[i,i+1][i+2,i+3]$ is an odd-parity state of the reflection.
Among all two-dimer states with $S_{\rm{tot}}^z=0$, there are two odd-parity states of the reflection,
$[i,i+1](i+2,i+3)$ and $(i,i+1)[i+2,i+3]$ with $(i,j)=\left\{ \left| \uparrow_i\downarrow_j\right> + \left| \downarrow_i\uparrow_j \right> \right\}/\sqrt{2}$.
However, these two states have odd parity under the spin inversion. Thus, we obtain  $(h_{i,i+2} - h_{i+1,i+3})[i,i+1][i+2,i+3]=0$,
which shows that $\Phi_1(L)$ and $\Phi_2(L)$ are eigenstates of $H_{\alpha=1}$, even with $\delta \neq 0$.
The energy eigenvalue of these singlet dimer states is given by
\begin{eqnarray}
E_{\mathrm{dimer}}=-\frac{\Delta + 2}{8}L.
\label{eq:5}
\end{eqnarray}
In order to prove that the dimer singlet state is a ground state, we rewrite the Hamiltonian in eq.~(\ref{eq:4}) as the sum of spin plaquette parts:
\begin{eqnarray}
   H_{\alpha=1} = \sum^{L/2}_{l=1}h^{\rm{plaq}}_l,
\label{eq:6}
\end{eqnarray}
with
\begin{eqnarray}
   h^{\rm{plaq}}_l= &&\hspace{-7mm} h_{2l,2l+1}+(1/2-\delta)(h_{2l-1,2l}+h_{2l-1,2l+1}) \nonumber \\
             &&\hspace{-7mm}+(1/2+\delta)(h_{2l+1,2l+2}+h_{2l,2l+2}).
\label{eq:7}
\end{eqnarray}
The minimum eigenvalue of $h^{\rm{plaq}}_l$ is $-(2+\Delta)/4$ for $\Delta \geq -1/2$ and $3\Delta/4$ otherwise.
Therefore, the ground-state energy $E_g$ of $H_{\alpha=1}$ satisfies
\begin{eqnarray}
   E_g\ge\left\{
   \begin{array}{cl}
      -\frac{2+\Delta}{8}L & \mbox{for $\Delta \geq -1/2$} \\
      \frac{3\Delta}{8}L & \mbox{for $\Delta < -1/2$}
   \end{array}
   \right..
\label{eq:8}
\end{eqnarray}
The lower bound of $E_g$ {for $\Delta \geq -1/2$ agrees with $E_{\mathrm{dimer}}$, 
which proves that the dimer singlet state is a ground state.
For $\Delta < -1/2$, the ferromagnetic state, whose energy is $E_{\mathrm{ferro}}=\frac{3\Delta}{8}L$, is a ground state.

We turn to the general case with $\alpha\neq 1$.
To obtain insight into the ground-state phase diagram, we bosonize the Hamiltonian in eq.~(\ref{eq:1}), which leads to
\begin{eqnarray}
H  = \frac{a}{2\pi}\int dx \left[ A(\de_x\phi)^2 + B(\pi\Pi)^2 \right]
   - C\int\frac{\cos4\phi}{(2\pi a)^2},
\label{eq:9}
\end{eqnarray}
where $a$ is a lattice constant. The coefficients $A$, $B$, and $C$ are, in general, functions of $\Delta$, $\alpha$ and $\delta$.
In the vicinity of $\Delta=0$ and $\alpha=0$, the explicit forms of $A$, $B$, and $C$ are
\begin{subequations}
\begin{eqnarray}
     A&&\hspace{-7mm}=1 + \frac{3\Delta}{\pi} +  \frac{\alpha(6 + \Delta)}{2\pi},  \\
\label{eq:10a}
     B&&\hspace{-7mm}=1 - \frac{\Delta}{\pi} - \frac{\alpha(2 - \Delta)}{2\pi},    \\
\label{eq:10b}
     C&&\hspace{-7mm}=a\left(\Delta - \frac{\alpha(2 + \Delta)}{2}\right).
\label{eq:10c}
\end{eqnarray}
\label{eq:10}
\end{subequations}
Note that $A$, $B$, and $C$ do not depend on $\delta$, which is because $\delta\sum_{i\in \mathrm{even}}h_{i,i+2}$
and $-\delta\sum_{i\in \mathrm{odd}}h_{i,i+2}$ in eq.~(\ref{eq:1}) cancel each other out in the long-wavelength limit.
This fact suggests that the phase boundaries in the $\alpha-\Delta$ plane do not change 
when $\delta$ varies from the railroad-trestle point $\delta=0$ to the sawtooth point $\delta=1/2$.

At the Heisenberg point, $\Delta=1$, the phase boundary between the dimer and N\'{e}el phases 
has already been determined for both the railroad-trestle and sawtooth models by using the LS method.
Previous studies estimated $\alpha^{\rm dimer/\mathrm{\Neel}}_{\rm c}(\Delta=1,\delta=0)=0.4822$ for the railroad-trestle model~\cite{okamoto} and 
$\alpha^{\rm dimer/\mathrm{\Neel}}_{\rm c}(\Delta=1,\delta=1/2)=0.4874$ for the sawtooth model,\cite{blundell} which indicates that
the relation $\alpha^{\rm dimer/\mathrm{\Neel}}_{\rm c}(\Delta=1,\delta=0) = \alpha^{\rm dimer/\mathrm{\Neel}}_{\rm c}(\Delta=1,\delta=1/2)$ suggested by the bosonization method approximately holds.

In this paper, we calculate the ground-state phase diagram of the asymmetric railroad-trestle model by using the LS method and study 
how the phase boundaries in the $\alpha-\Delta$ plane depend on $\delta$.
As a result, we find that the $\delta$-dependence of the dimer-N\'{e}el critical line is small, but that of the dimer-SF critical line is large.

%%%%%%%%%%%%%%%%%%%%%%%%%%%%%%%%%%%%%%%%%%%%%%%%%%%%
\section{LS Method}%%%%%%%%%%%%%%%%%%%%%%%%%%%%%%%%%%%%%%%%%%
%%%%%%%%%%%%%%%%%%%%%%%%%%%%%%%%%%%%%%%%%%%%%%%%%%%%

We now describe our calculations based on the LS method.\cite{okamoto,nomura0,nomura1,nomura2}
In Fig.~\ref{fig:LS}, we show the excitation energies of N\'{e}el, dimer, and doublet excitations as functions of $\alpha$ in a finite-size cluster with $L=20$.
The intersection, $\alpha_{\rm c}(L; \Delta=0.5,\delta=0)$, between the dimer and doublet excitation energies in Fig.~\ref{fig:LS}(a) is interpreted 
as a dimer-SF transition point in the $L=20$ cluster. Also, the intersection, $\alpha_{\rm c}(L;\;\Delta=2,\delta=1/2)$, 
between the dimer and N\'{e}el excitation energies in Fig.~\ref{fig:LS}(a) is interpreted as a dimer-N\'{e}el transition point.
The size dependence of $\alpha_{\rm c}(L; \Delta,\delta)$ is expected to be~\cite{okamoto,nomura0}
\begin{equation}
     \alpha_{\rm c}(L; \Delta,\delta) = \alpha_{\rm c}(\infty; \Delta,\delta) + \mathrm{Const.}\times L^{-2}.
\label{eq:11}
\end{equation}
Our calculated data of $\alpha_{\rm c}(L; \Delta,\delta)$ with $L=12$, 16, $\cdots$, 28 are plotted against $L^{-2}$ in the insets in Fig.~\ref{fig:LS}.
We can confirm from this figure that our data follow the size dependence in eq.~(\ref{eq:11}) and that the extrapolation to $L=\infty$ is satisfactory.

Next, we calculate the central charge $c$ for verifying our calculations.
According to conformal field theory, we can estimate $c$ using the following expression for the ground-state energy:
\begin{equation}
E_0(L) = \epsilon_0 L - \frac{\pi v c}{6L},
\label{eq:12}
\end{equation}
with the spin-wave velocity $v = L\Delta E(k = 2\pi/L)/2\pi$.
Figure~\ref{fig:cc}(a) shows the $\delta$-dependence of $c$ for $(\delta,\Delta)=(0,0)$ and $(0.5,2,5)$.
As expected, it is found that $c=1$ in the SF phase ($\alpha<0.6474$) for $(\delta,\Delta)=(0,0)$ and at the dimer-$\Neel$ transition point ($\alpha=0.7048$) for $(\delta,\Delta)=(0.5,2.5)$.
We also calculate $c$ along the dimer-SF and dimer-$\Neel$ critical lines, where $c=1$ is expected.
The result is shown in Fig.~\ref{fig:cc}(b).
Our result is consistent with $c=1$, although a small error is observed.
%
%----------------------------------------------------
\begin{figure}[h]
\begin{center}
\begin{minipage}[t]{0.24\textwidth}
%\begin{minipage}[t]{0.3\textwidth}
\includegraphics[width=\linewidth]{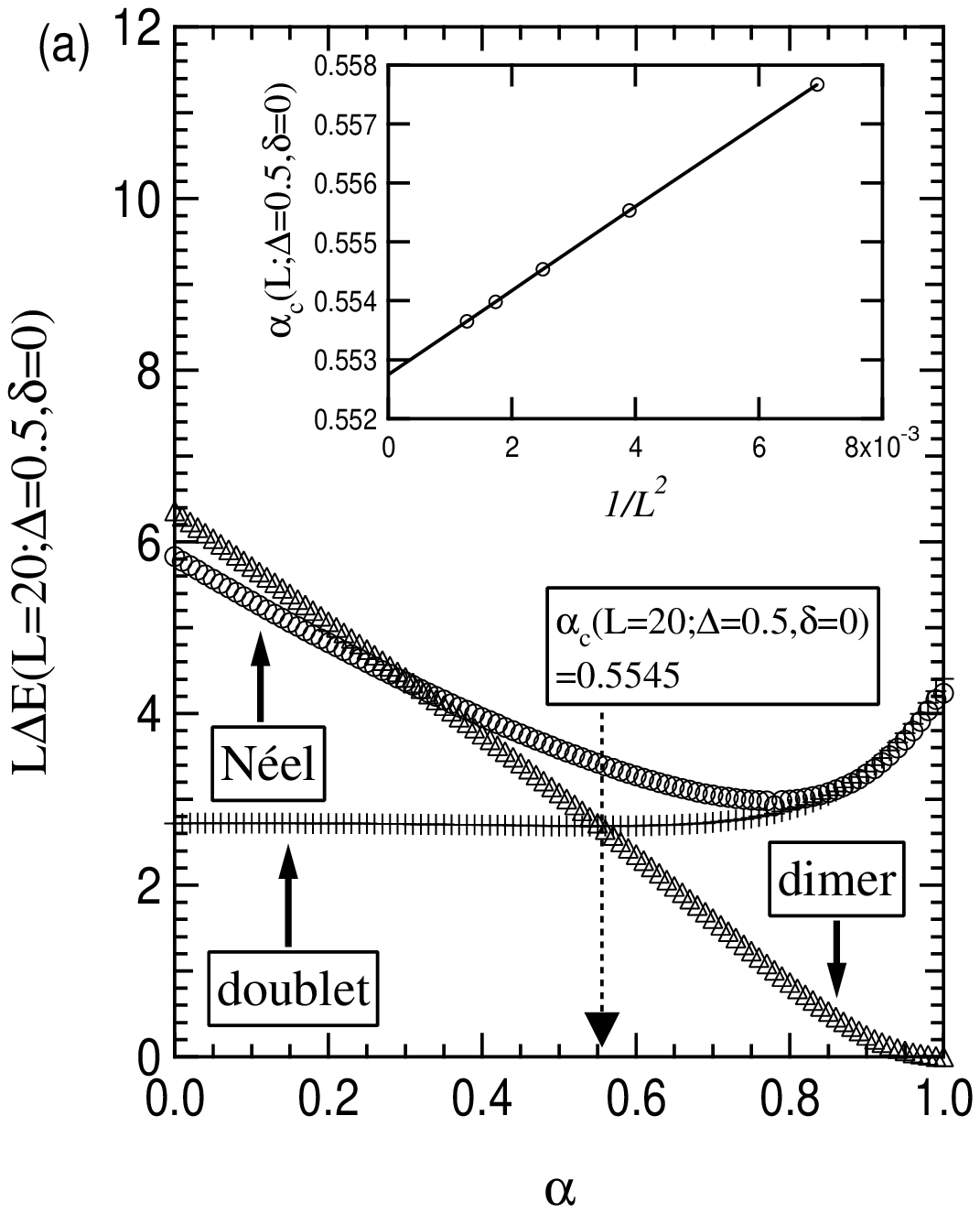}
\end{minipage}
\begin{minipage}[t]{0.24\textwidth}
%\begin{minipage}[t]{0.3\textwidth}
\includegraphics[width=\linewidth]{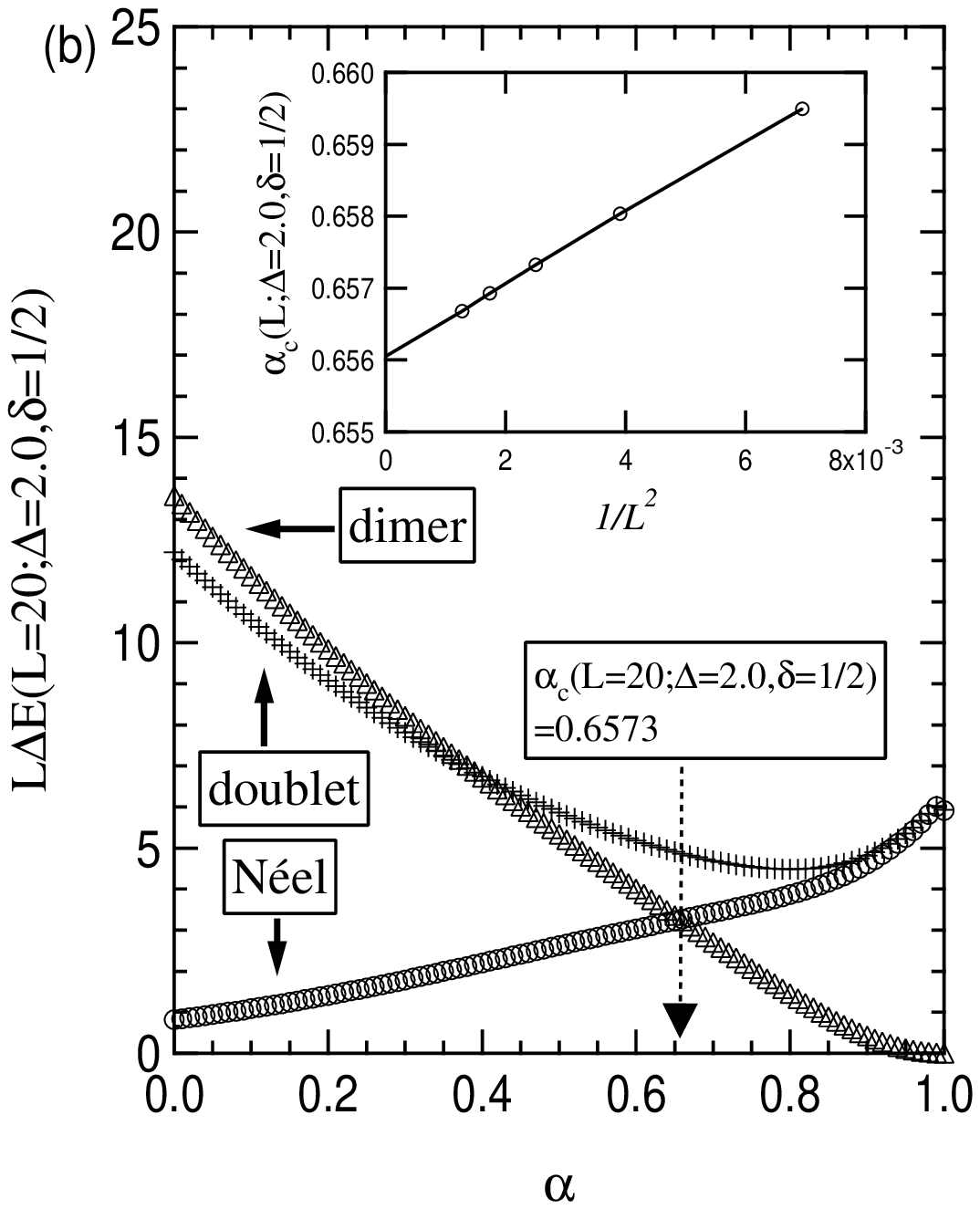}
\end{minipage}
\end{center}
\caption{Excitation energies of N\'{e}el, dimer, and doublet excitations as functions of $\alpha$ for (a) $\delta=0$, $\Delta=0.5$
and (b) $\delta=1/2$, $\Delta=2$, where the system size is $L=20$.}
\label{fig:LS}
\end{figure}
%----------------------------------------------------
%----------------------------------------------------
\begin{figure}[h]
\begin{center}
\begin{minipage}[t]{0.24\textwidth}
%\begin{minipage}[t]{0.3\textwidth}
\includegraphics[width=\linewidth]{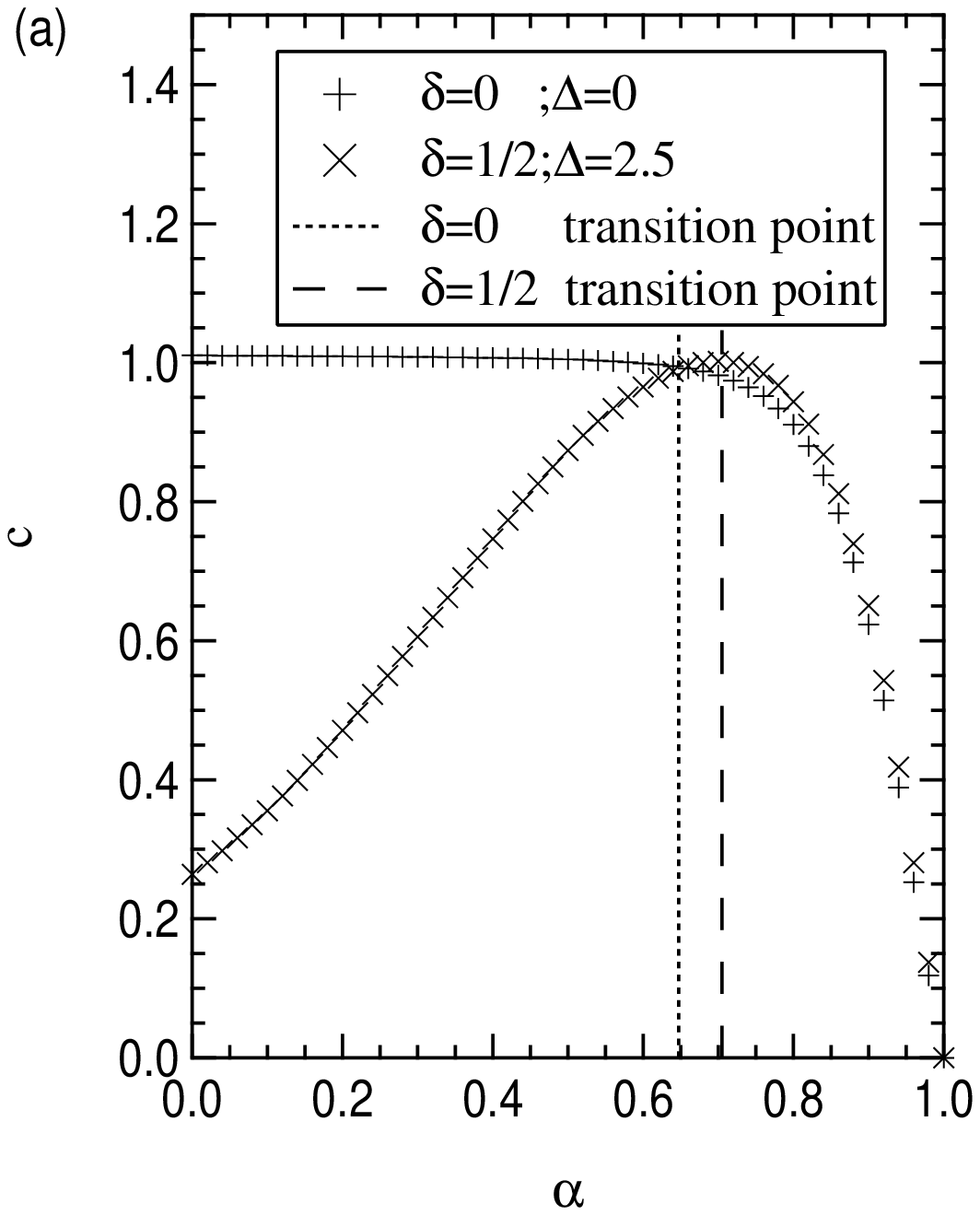}
\end{minipage}
\begin{minipage}[t]{0.24\textwidth}
%\begin{minipage}[t]{0.3\textwidth}
\includegraphics[width=\linewidth]{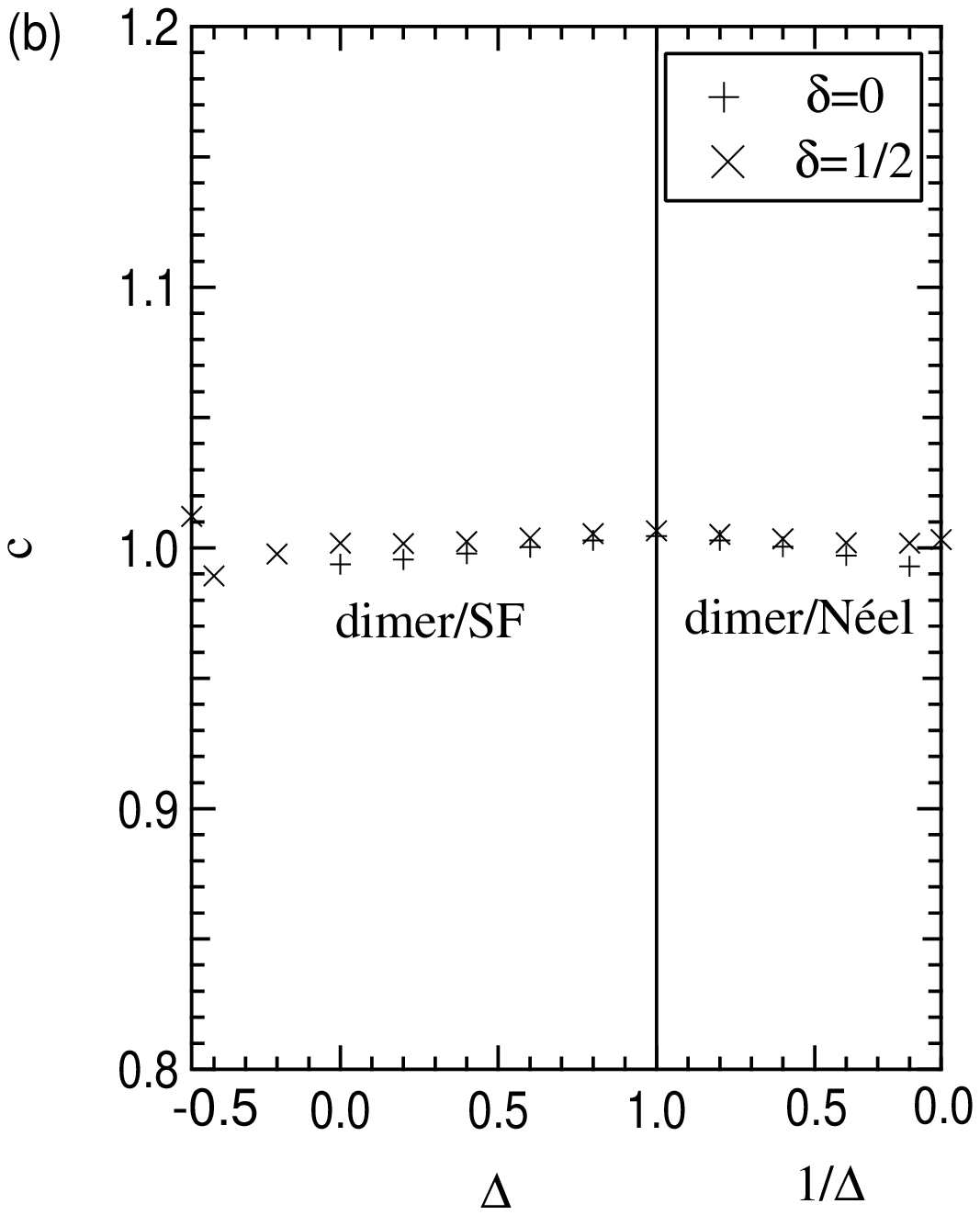}
\end{minipage}
\end{center}
\caption{(a) Plot of the central charge $c$ against $\alpha$ for fixed values of $\delta$ and $\Delta$. (b) Central charge $c$ on the critical lines as a function of $\Delta$.}
\label{fig:cc}
\end{figure}
%----------------------------------------------------

%%%%%%%%%%%%%%%%%%%%%%%%%%%%%%%%%%%%%%%%%%%%%%%%%%%%
\section{Ground-State Phase Diagram}%%%%%%%%%%%%%%%%%%%%%%%%%%%%%%
%%%%%%%%%%%%%%%%%%%%%%%%%%%%%%%%%%%%%%%%%%%%%%%%%%%%

Putting together all our calculated data, we obtain the phase diagram in the $\alpha-\Delta$ plane shown in Fig.~\ref{fig:phase}, 
where the phase boundaries for $\delta=0$, $1/4$ and $1/2$ are presented. 
Our result for $\delta=0$ reproduces the ground-state phase diagram of the railroad-trestle model 
previously obtained by Nomura and Okamoto~\cite{nomura0} and Hirata and Nomura.~\cite{hirata}
We find in Fig.~\ref{fig:phase} that the dimer-N\'{e}el critical line and the first-order phase transition line to the ferromagnetic state are almost independent of $\delta$. 
On the other hand, the dimer-SF critical line depends on $\delta$, and this dependence is pronounced at approximately $\Delta=-0.5$.
As a result, the region of direct transition between the dimer and ferromagnetic states~\cite{hirata} becomes narrower when $\delta$ increases,
and this region vanishes at the sawtooth point $\delta=1/2$.
The pronounced $\delta$-dependence of the dimer-SF critical line contradicts the naive expectation based on the bosonization.

%--------------------------------------------------------------------
\begin{figure}[h]
\begin{center}
\includegraphics[width=0.8\linewidth]{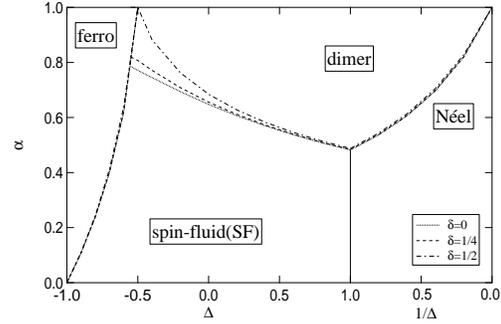}
\end{center}
\caption{The ground-state phase diagram in the $\alpha-\Delta$ plane for $\delta=0$, $1/4$ and $1/2$.}
\label{fig:phase}
\end{figure}
%--------------------------------------------------------------------

To see the $\delta$-dependence of the transition lines in more detail, we define 
\begin{equation}
   \alpha'(\Delta,\delta)=\alpha_{\rm c}(\Delta,\delta)-\alpha_{\rm c}(\Delta,0)
\label{eq:13}
\end{equation}
and show $\alpha'(\Delta,1/4)$ and $\alpha'(\Delta,1/2)$ in Fig.~\ref{fig:deviation}.
We find that $\alpha'(\Delta,1/2)$ for the dimer-N\'{e}el and ferro-dimer transitions is at most $\sim 10^{-2}$.
For the dimer-SF transition, $\alpha'(\Delta,1/2)$ increases as $\Delta$ decreases, and it amounts to $\sim 10^{-1}$ at approximately $\Delta=-0.5$.
Roughly speaking, the $\delta$-dependence of the dimer-SF transition is about ten times as large as that of the dimer-N\'{e}el and ferro-SF transitions.
% The broken lines denote $\alpha'_{\rm c}(\Delta,\delta)$ for the $L=4$ cluster, 

%--------------------------------------------------------------------
\begin{figure}[h]
\begin{center}
\includegraphics[width=0.8\linewidth]{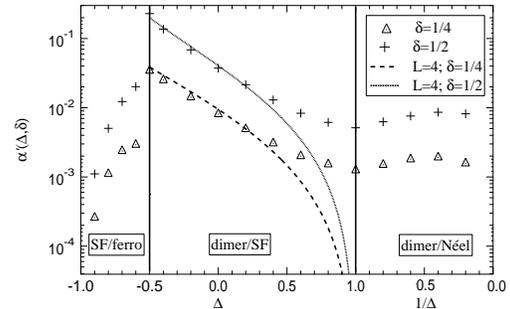}
\end{center}
\caption{Plot of $\alpha'(\Delta,\delta)=\alpha_{\rm c}(\Delta,\delta)-\alpha_{\rm c}(\Delta,0)$
as a function of $\Delta$ for $\delta=1/2$ and 1/4. The markers are the results in the thermodynamic limit and the lines are those for $L=4$.}
\label{fig:deviation}
\end{figure}
%--------------------------------------------------------------------

%%%%%%%%%%%%%%%%%%%%%%%%%%%%%%%%%%%%%%%%%%%%%%%%%%%%
\section{Discussion} %%%%%%%%%%%%%%%%%%%%%%%%%%%%%%%%%%%%%%%%
%%%%%%%%%%%%%%%%%%%%%%%%%%%%%%%%%%%%%%%%%%%%%%%%%%%%

In this section, we study the  eigenvectors for the ground state, dimer excitation, N\'{e}el excitation, and doublet excitation to discuss 
why the $\delta$-dependence of the dimer-SF transition line is much larger than those of the dimer-N\'{e}el and ferro-SF transition lines.

We begin by representing $H$ graphically and defining a reflection operation $\hat{I}$ and a translation operation $\hat{t}$, as shown in Fig.~\ref{fig:L4n}.
Note that $\hat{I}$ and $\hat{t}$ do not change $H$ only when $\delta=0$, but change $H$ when $\delta\neq 0$.
We also define the spin inversion operation as $\hat{T}$. Hereafter, we denote the eigenvalues of  $\hat{T}$, $\hat{t}$, and $\hat{I}$ as
$T$, $t=e^{i k}$, and $I$, respectively.

%%%%%%%%%%%%%%%%%%%%%%%%%%%%
\begin{figure}[h]
\begin{center}
\includegraphics[width=0.6\linewidth]{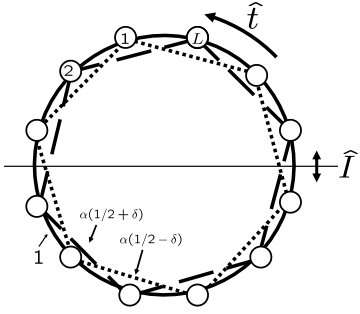}
\end{center}
\caption{Another graphical representation of the Hamiltonian
in eq.~(\ref{eq:1}). We denote the translation operation as $\hat{t}$
and the reflection operation about the horizontal thin solid line as $\hat{I}$.}
\label{fig:L4n}
\end{figure}
%%%%%%%%%%%%%%%%%%%%%%%%%%%%

When $\delta=0$, it is known that the ground state is in a $S_{\rm{tot}}^z=0$ subspace with $(T,k,I)=(1,0,1)$,
the dimer excitation is in the subspace with $(1,\pi,1)$, and the N\'{e}el excitation is in the subspace with $(-1,\pi,-1)$.~\cite{nomura0}
Operation of $\delta\sum_i (-1)^i h_{i,i+2}$ to basis functions with $(T,k,I)=(1,0,1)$, $(1,\pi,1)$ or $(-1,\pi,-1)$
yields basis functions with $(T,k,I)=(1,\pi,-1)$, $(1,0,-1)$ or $(-1,0,1)$, respectively. 
Thus, for $\delta \geq 0$, the ground-state wave function,  $|\Psi_{\mathrm{gs}}\rangle$,  
the wave function of the dimer excitation, $|\Psi_{\mathrm{dimer}}\rangle$, and that of the N\'{e}el excitation, $|\Psi_{\mathrm{\Neel}}\rangle$,
are written as
\begin{subequations}
\begin{eqnarray}
   |\Phi_{\mathrm{gs}}\rangle&&\hspace{-7mm}= |\phi_{1,0,1}\rangle\cos\theta_{\mathrm{gs}} + |\phi_{1,\pi,-1}\rangle\sin\theta_{\mathrm{gs}},
   \label{eq:14a}\\
   |\Phi_{\mathrm{dimer}}\rangle&&\hspace{-7mm}= |\phi_{1,\pi,1}\rangle\cos\theta_{\mathrm{dimer}} + |\phi_{1,0,-1}\rangle\sin\theta_{\mathrm{dimer}},\;\;\;\;\;\;
   \label{eq:14b}\\
   |\Phi_{\mathrm{\Neel}}\rangle&&\hspace{-7mm}= |\phi_{-1,\pi,-1}\rangle\cos\theta_{\mathrm{\Neel}} + |\phi_{-1,0,1}\rangle\sin\theta_{\mathrm{\Neel}},\;\;\;\;\;\;
   \label{eq:14c}
\end{eqnarray}
\label{eq:14}
\end{subequations}
\noindent
where $|\phi_{T,k,I}\rangle$ represents unit vector in the $(T,k,I)$ subspace of $|\Phi_{\mathrm{gs}}\rangle$, $|\Phi_{\mathrm{dimer}}\rangle$ or $|\Phi_{\mathrm{\Neel}}\rangle$,
respectively, and we choose $0 \leq \theta \leq \pi/2$. (We hereafter call $\theta$ the ``mixing parameter".)
%%%%%%%%%%%%%%%%%%%%%%%%%%%%%%%%%%%%
\begin{table}[b]
\centering
\caption{Symmetrized basis functions with $S^z_{\rm tot}=0$ for the $L=4$ cluster. The eigenvalues of $\hat{T}$,
$\hat{t}$ and $\hat{I}$ are denoted as $T$, $e^{ik}$ and $I$, respectively.}
\vspace{1mm}
\label{table:basis1}
\begin{tabular}{@{\hspace{\tabcolsep}\extracolsep{\fill}}c|l}
\hline
$(T, k, I)$ & \hspace{1cm}basis function \\
\hline
$(1,0,1)$                           & $(|s s\rangle+|t_0 t_0\rangle)/\sqrt{2}$  \\
                                          & $(|s s\rangle-|t_0 t_0\rangle-|t_+ t_-\rangle-|t_- t_+\rangle)/2$  \\
\hline
$(1,\pi,1)$                        & $(|s s\rangle-|t_0 t_0\rangle+|t_+ t_-\rangle+|t_- t_+\rangle)/2$  \\
\hline
$(-1,\pi,-1)$                     & $(|s t_0\rangle+|t_0 s \rangle)/\sqrt{2}$  \\
\hline
$(-1,\frac{\pi}{2},\ast)$  & $(|s t_0\rangle-|t_0 s \rangle-i |t_+ t_-\rangle+i|t_- t_+\rangle)/2$  \\
\hline
$(-1,-\frac{\pi}{2},\ast)$ & $(|s t_0\rangle-|t_0 s \rangle+i |t_+ t_-\rangle-i|t_- t_+\rangle)/2$  \\
\hline
\end{tabular}
\end{table}
%%%%%%%%%%%%%%%%%%%%%%%%%%%%%%%%%%%%
%%%%%%%%%%%%%%%%%%%%%%%%%%%%%%%%%%%%
\begin{table}[t]
\centering
\caption{Symmetrized basis functions with total $S^z_{\rm tot}=1$, 2 for the $L=4$ cluster.}
\vspace{1mm}
\label{table:basis2}
\begin{tabular}{@{\hspace{\tabcolsep}\extracolsep{\fill}}c|l}
\hline
$(S^z_{\rm tot},k,I)$ & \hspace{1cm}basis function \\
\hline
$(1,0,1)$                         & $(|t_0t_+\rangle+|t_+t_0\rangle)/\sqrt{2}$  \\
\hline
$(1,\pi,-1)$                     & $(|s t_+\rangle+|t_+ s\rangle)/\sqrt{2}$  \\
\hline
$(1,\frac{\pi}{2},\ast)$  & $(|s t_+\rangle-|t_+ s\rangle-i|t_0t_+\rangle+i|t_+t_0\rangle)/2$  \\
\hline
$(1,-\frac{\pi}{2},\ast)$ & $(|s t_+\rangle-|t_+ s\rangle+i|t_0t_+\rangle-i|t_+t_0\rangle)/2$  \\
\hline
$(2,0,1)$ & $|t_+t_+\rangle$  \\
\hline
\end{tabular}
\end{table}
%%%%%%%%%%%%%%%%%%%%%%%%%%%%%%%%%%%%
%%%%%%%%%%%%%%%%%%%%%%%%%%%%%%%%%%%%
\begin{table}[t]
\caption{Total numbers of basis functions in the subspaces relevant to the determination of transition lines.}
\vspace{1mm}
\begin{center}
\begin{tabular}{c|c|c|c|c|c}\hline
state   &$(T~\mathrm{or}~S^z_{\mathrm{tot}},k,I)$ & $L=4$ & $8$ & $12$ & $16$     \\ \hline
ground  &$(1,0,1)$     & 2   & 7 & 35 & 257     \\ 
 state  &$(1,\pi,-1)$  & 0   & 1 & 13 & 175     \\ \hline
dimer   &$(1,\pi,1)$   & 1   & 5 & 29 & 239     \\ 
excitation &$(1,0,-1)$    & 0   & 0 & 9  & 158     \\ \hline
$\Neel$ &$(-1,\pi,-1)$ & 1   & 4 & 27 & 230     \\ 
 excitation &$(-1,0,1)$    & 0   & 1 & 15 & 183     \\ \hline
doublet &$(1,\pi,-1)$  & 1   & 5 & 38 & 375     \\ 
 excitation &$(1,0,1)$     & 1   & 5 & 38 & 375     \\ \hline
\end{tabular}
\end{center}
\label{table:basisnumber}
\end{table}
%%%%%%%%%%%%%%%%%%%%%%%%%%%%%%%%%%%%
For the doublet excitation,
it is known that the symmetry of the doublet excitation is characterized by $(S_{\rm{tot}}^z,k,I)=(1,\pi,-1)$ when $\delta=0$.
Operation of $\delta\sum_i (-1)^i h_{i,i+2}$ to those basis functions yields basis functions with $(1,0,1)$.
We can write the wave function of the doublet excitation for $\delta\geq 0$ as
\begin{equation}
     |\Phi_{\mathrm{doublet}}\rangle=|\varphi_{1,\pi,-1}\rangle \cos \theta_{\mathrm{doublet}}+|\varphi_{1,0,1}\rangle \sin \theta_{\mathrm{doublet}},
     \label{eq:15}
\end{equation}
where $|\varphi_{S_{\rm{tot}}^z,k,I}\rangle$ is unit vector in the $(S_{\rm{tot}}^z,k,I)$ subspace of $|\Phi_{\mathrm{doublet}}\rangle$.

Here we consider the case of $L=4$.
It is known that LS analysis gives good results even for small systems;
thus, we expect that the study of the $L=4$ cluster is a good starting point for understanding the $\delta$-dependence of the transition lines.
The symmetrized basis functions are tabulated in Tables~\ref{table:basis1} and \ref{table:basis2},
where the basis functions are expressed by dimer states, 
$|s\rangle=\frac{1}{\sqrt{2}}(|\uparrow\downarrow\rangle-|\downarrow\uparrow\rangle)$, 
$|t_0\rangle=\frac{1}{\sqrt{2}}(|\uparrow\downarrow\rangle+|\downarrow\uparrow\rangle)$, $|t_+\rangle=|\uparrow\uparrow\rangle$,
and $|t_-\rangle=|\downarrow\downarrow\rangle$. 
For example, a four-spin state $|s t_0\rangle$ means that two spins on the first and second sites
are in the singlet dimer state and  those on the third and fourth sites are in the triplet dimer state with $S^z_{\rm{tot}}=0$.
Note that there are no basis functions with $(T,k,I)=(1,\pi,-1),(1,0,-1),$ or $(-1,0,1)$. (See also Table~\ref{table:basisnumber}, 
where the total numbers of basis functions in each of the subspaces are listed.)
Therefore, we obtain
\begin{equation}
   \theta_{\mathrm{gs}} =  \theta_{\mathrm{dimer}} = \theta_{\mathrm{\Neel}} = 0,
\label{eq:16}
\end{equation}
which means that the energy eigenvalues of the ground state, dimer excitation, and N\'{e}el excitation do not depend on $\delta$,
and thus $\alpha_{\rm c}^{\rm{dimer}/\Neel}$ and $\alpha_{\rm c}^{\rm{SF}/\rm{ferro}}$ are completely independent of $\delta$ in the $L=4$ cluster.
For the doublet excitation, however, there is a basis function in the $(S^z_{\rm{tot}}=1, k=0, I=1)$ subspace.
Therefore, $\theta_{\mathrm{doublet}}$ does not vanish. Using the basis functions in Table~\ref{table:basis2},
we obtain
\begin{equation}
     \theta_{\mathrm{doublet}} =\arctan \left(\frac{\sqrt{1+[\alpha\delta(1-\Delta)]^2}-1}{\alpha\delta(1-\Delta)}\right),
     \label{eq:17}
\end{equation}
which means that $\alpha_{\rm c}^{\rm dimer/SF}$ depends on $\delta$.
An explicit calculation gives
\begin{equation}
     \alpha_{\rm c}^{\rm dimer/SF} = \frac{2}{\sqrt{(3+\Delta)^2-4\delta^2(\Delta-1)^2}}.
     \label{eq:18}
\end{equation}
Using the results obtained above, $\alpha'(\Delta,\delta)$, which is defined in eq.~(\ref{eq:13}), for the $L=4$ cluster is calculated as
\begin{equation}
     \alpha'_{L=4}=\left\{
     \begin{array}{cc}
     \frac{2}{3+\Delta}\left(\frac{1}{\sqrt{1-\left[\frac{2\delta(1-\Delta)}{3+\Delta}\right]^2}}-1\right)  & \mbox{for dimer/SF} \\
     0 & \mbox{otherwise} \\
     \end{array}\right..
     \label{eq:19}
\end{equation}
The present analysis of the $L=4$ cluster shows that the pronounced $\delta$-dependence of $\alpha'(\Delta,\delta)$ for only the dimer-SF transition
is because only the mixing parameter $\theta_{\mathrm{doublet}}$ for the doublet excitation can become finite.

In Fig.~\ref{fig:deviation}, we show $\alpha'_{L=4}(\Delta,\delta)$ for $\delta=1/4$ (dashed line) and 1/2 (solid line). 
We find that $\alpha'_{L=4}(\Delta,\delta)$ shows a similar characteristic behavior to $\alpha'_{L=\infty}(\Delta,\delta)$.
In particular, $\alpha'_{L=4}(\Delta,\delta)$ agrees well with $\alpha'_{L=\infty}(\Delta,\delta)$ for $-0.5<\Delta<\sim 0.5$.
However, $\alpha'_{L=\infty}(\Delta,\delta)$ for N\'{e}el-dimer and ferro-SF transitions is small but finite, although $\alpha'_{L=4}(\Delta,\delta)=0$. 
As shown in Table~\ref{table:basisnumber}, basis functions with $(T,k,I)=(1,\pi,-1),(1,0,-1)$, and $(-1,0,1)$ appear for larger clusters,
so it is possible to be $\theta_{\mathrm{gs}}$, $\theta_{\mathrm{dimer}}$, $\theta_{\mathrm{\Neel}}\neq 0$.
Even in such larger clusters, we expect that these values remain smaller than that of the doublet excitation.
We confirm this fact numerically for a larger cluster ($L=12$) below.

We now study the mixing parameters for the $L=12$ cluster.
Here, $(\Delta, \alpha)$ is fixed to be $(2.0, 0.7)$ and $(-0.5, 0.9)$, where the former and latter are chosen to be near
the dimer-N\'{e}el and  dimer-SF transition lines, respectively.
We calculate the mixing parameters as functions of $\delta$.
The result for $(\Delta, \alpha)=(2.0, 0.7)$ is shown in Fig.~\ref{fig:theta}(a).
We find that all the mixing parameters are finite for $\delta>0$ and these values increase as $\delta$ increases.
Although $\theta_{\mathrm{\Neel}}$ is comparable to $\theta_{\mathrm{doublet}}$, these values are very small, $0.04\times \pi/2$ at the most.
The smallness of the mixing parameters results in the dimer-N\'{e}el transition line being almost independent of $\delta$.
Next, we show the result for $(\Delta, \alpha)=(-0.5, 0.9)$ in Fig.~\ref{fig:theta}(b).
As expected, we find that $\theta_{\mathrm{doublet}}$ is much larger than the other values, and it amounts to $0.17\times \pi/2$ at $\delta=0.5$,
which leads to the pronounced $\delta$-dependence of the dimer-SF transition line.
Thus, we conclude that the feature extracted from the $L=4$ cluster analysis, the $\delta$-dependence of the dimer-SF transition line due to the mixing of the doublet excitation,
also holds for larger clusters.
\begin{figure}[h]
\begin{center}
%\subfigure[ ]{\includegraphics[width=0.7\linewidth]{sin_1.eps}\label{sin_1}}
%\subfigure[ ]{\includegraphics[width=0.7\linewidth]{sin_2.eps}\label{sin_2}}
\includegraphics[width=0.7\linewidth]{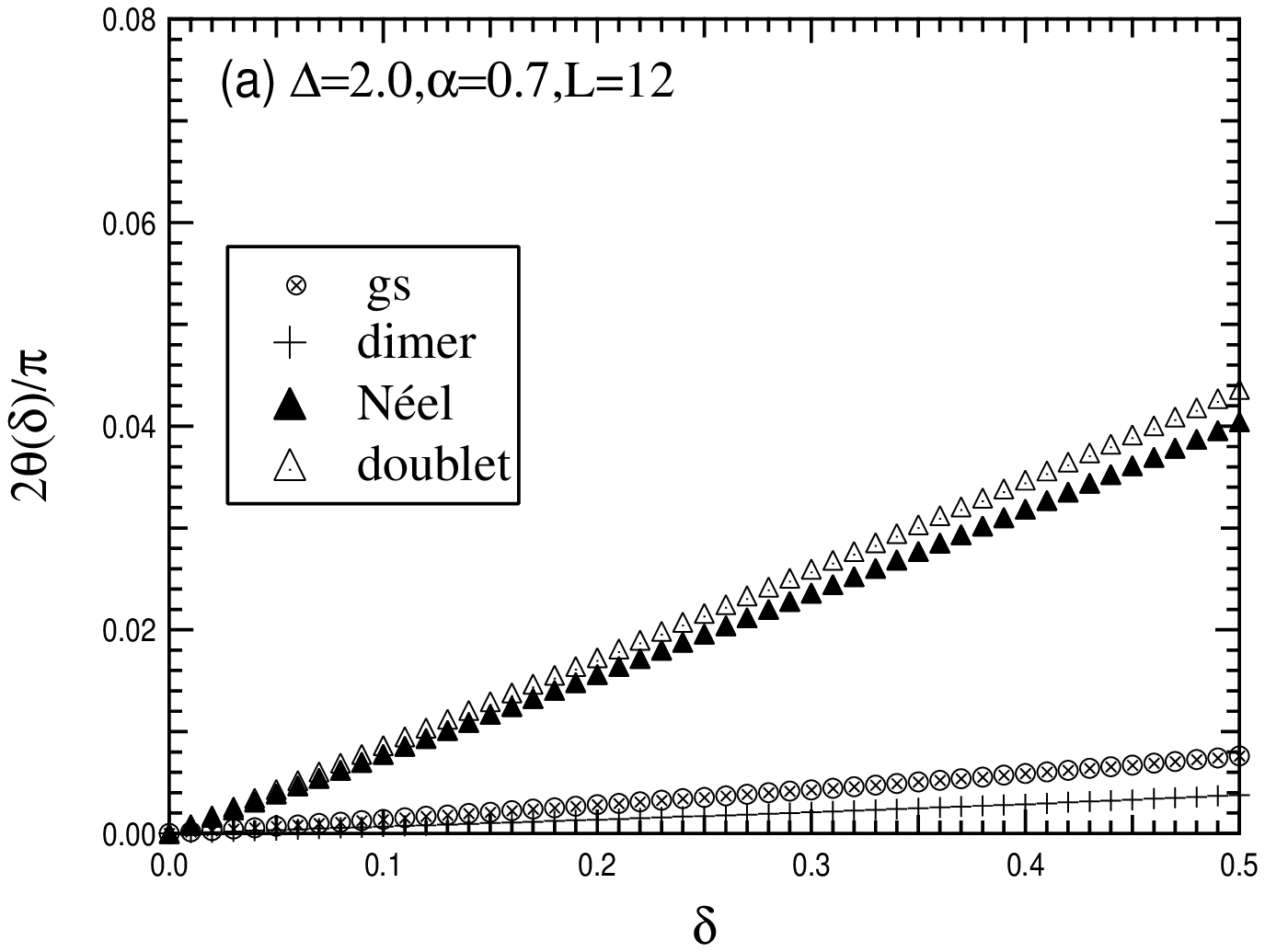}
\includegraphics[width=0.7\linewidth]{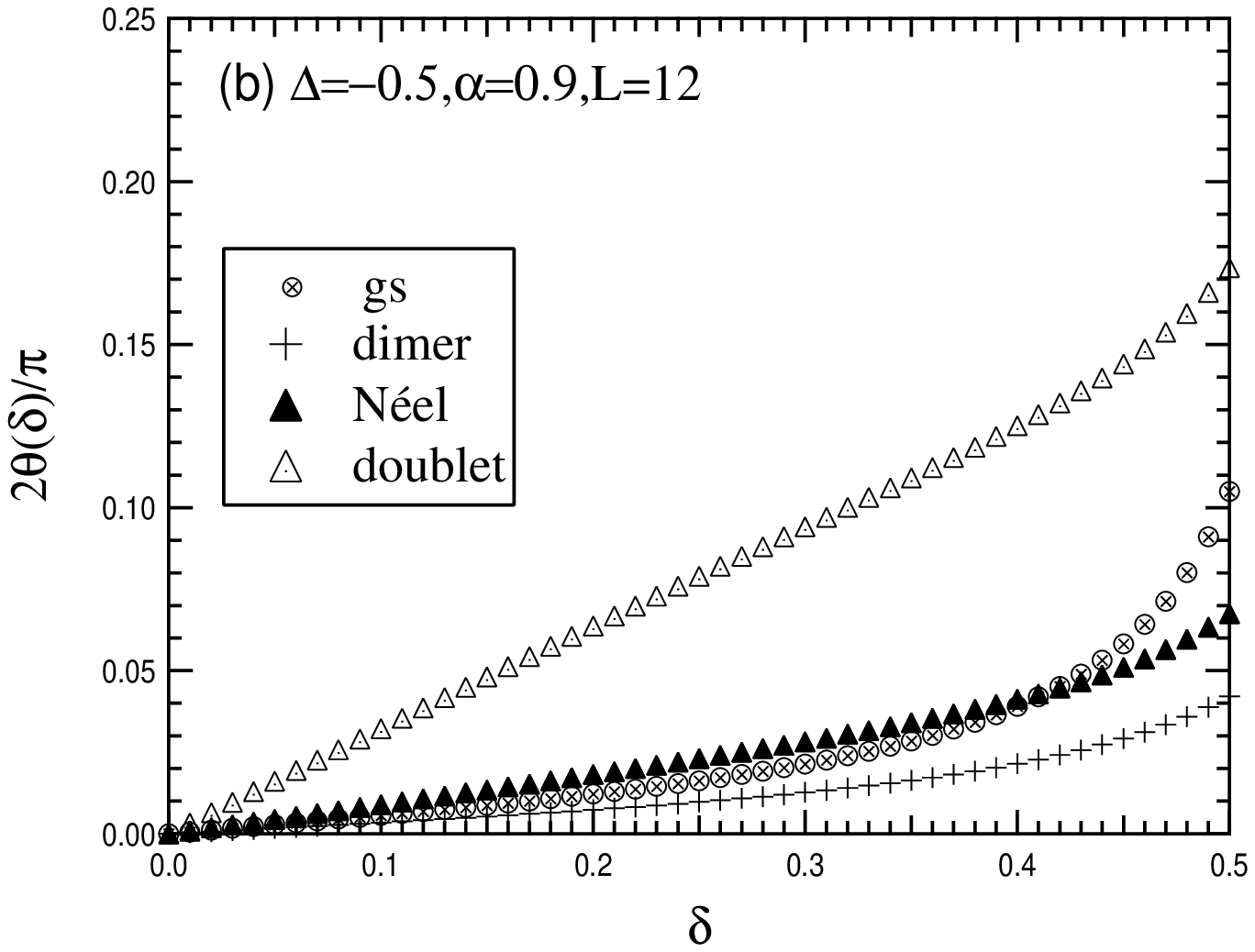}
\end{center}
\caption{Mixing parameter $\theta$ against $\delta$ for (a) $\Delta=2.0$, $\alpha=0.7$ and
(b) $\Delta=-0.5$, $\alpha=0.9$. The system size is $L=12$.}
\label{fig:theta}
\end{figure}
%

%%%%%%%%%%%%%%%%%%%%%%%%%%%%%%%%%%%%%%%%%%%
\section{Summary}%%%%%%%%%%%%%%%%%%%%%%%%%%%%%%%%%
%%%%%%%%%%%%%%%%%%%%%%%%%%%%%%%%%%%%%%%%%%%

The ground-state phase diagram of the $XXZ$ model on the railroad-trestle lattice with asymmetric leg interactions has been investigated.
We have proven the existence of an exact singlet dimer ground state at $\alpha=1$, and have studied how the phase boundaries depend on the asymmetry parameter $\delta$.
The naive bosonization method predicted that the asymmetry does not affect the critical lines, but this prediction has turned out to be an oversimplification.
By using the LS method, we found that the dimer-N\'{e}el critical line and the ferro-SF first-order transition line are almost independent of $\delta$,
and the dimer-SF critical line depends on $\delta$, which is because the asymmetric interaction affects the doublet state,
but it does not affect the ground state, dimer excitation or N\'{e}el excitation significantly.
%We pointed out that the numbers of basis functions in symmetry subspaces at $\delta=0$ is a stable measure to know how the energy of each state depends on $\delta$.
%From this view point, we showed that the subspace to which the doublet excitation belongs has structure that is easy to be affected by the asymmetric interactions,
%but the subspaces to which the ground state, dimer excitation or N\'{e}el excitation belong do not. 
This may be interpreted as follows: 
the doublet excitation is a type of spin-wave state in the quasi long-range ordered state, and a spin wave, which is a local defect propagating in the quasi long-range ordered state,
can detect the short-wavelength part of the Hamiltonian, such as the asymmetric interaction.

After completion of this work, we became aware of studies by Sarkar and Sen~\cite{Sarkar}
and by Capriotti {\it et al.}~\cite{Capriotti} concerned with Heisenberg antiferromagnets on the asymmetric railroad-trestle lattice.
Sarkar and Sen produced an operator product expansion of the terms in the bosonized Hamiltonian
from the asymmetric part in leg interactions, and showed that these terms cancel each other.
Capriotti {\it et al.} numerically calculated the dimer/SF transition point as a function of
$\delta$ at the Heisenberg point, and they found that the $\delta$-dependence is very small.

\acknowledgements
We have used a part of the code provided by H. Nishimori in TITPACK Ver.~2.
We thank Professor D.~Sen for informing us of the two studies in refs' 18 and 19.

\end{document}